\documentclass[12pt]{article}
\usepackage[dvips]{graphics,lscape}
\usepackage{rotating,supertabular}
\usepackage{aasms4,amsmath,amsfonts,epsfig,psfig,float}

\def\eps@scaling{.5}
\def\epsscale#1{\gdef\eps@scaling{#1}}
\def\plotone#1{\centering \leavevmode
\epsfxsize=\eps@scaling\columnwidth \epsfbox{#1}}

\def\simlt{\lower.5ex\hbox{$\; \buildrel < \over \sim \;$}}
\def\simgt{\lower.5ex\hbox{$\; \buildrel > \over \sim \;$}}

\begin{document}
\setcounter{figure}{0}
\centerline{Submitted to the Editor of the Astrophysical Journal}
\bigskip
\title{A Magnetic Dynamo Origin For The Sub-mm Excess In Sgr A*}

\author{Fulvio Melia$^{\dag}$\altaffilmark{1}, Siming Liu$^*$, and Robert 
Coker$^{\dag\dag}$} 
\affil{$^{\dag}$Physics Department and Steward Observatory, The University 
of Arizona, Tucson, AZ 85721}
\affil{$^*$Physics Department, The University of Arizona, Tucson, AZ 85721}
\affil{$^{\dag\dag}$Department of Physics \& Astronomy, The University of Leeds, 
Leeds  LS2 9JT, UK}
\altaffiltext{1}{Sir Thomas Lyle Fellow and Miegunyah Fellow.}

\begin{abstract}
The sub-mm bump observed in the spectrum of Sgr A* appears to indicate
the existence of a compact emitting component within several Schwarzschild
radii, $r_S$, of the nucleus at the Galactic Center.  This is interesting in view
of the predicted circularized flow within $\sim 5-10\,r_S$, based on
detailed multi-dimensional hydrodynamic simulations of Bondi-Hoyle accretion
onto this unusual object.  In this paper, we examine the physics 
of magnetic field generation by a Keplerian dynamo subject to the conditions
pertaining to Sgr A*, and show that the sub-mm bump can be produced by
thermal synchrotron emission in this inner region.  This spectral feature
may therefore be taken as indirect evidence for the existence of this
circularization.  In addition, the self-Comptonization of the sub-mm
bump appears to produce an X-ray flux exceeding that due to bremsstrahlung
from this region, which may account for the X-ray counterpart to Sgr A*
discovered recently by {\it Chandra}.  However, the required accretion rate 
in the Keplerian flow is orders of magnitude smaller than that predicted by 
the Bondi-Hoyle simulations.  We speculate that rapid evaporation, in the form 
of a wind, may ensue from the heating associated with turbulent mixing of gas 
elements with large eccentricity as they settle down into a more or less circular 
(i.e., low eccentricity) trajectory.  The spectrum of Sgr A* longward of
$\sim 1-2$ mm may be generated outside of the Keplerian flow, where the 
gas is making a transition from a quasi-spherical infall into a
circularized pattern.
\end{abstract}
\keywords{accretion---black hole physics---hydrodynamics---Galaxy:
center---magnetic fields: dynamo---magnetohydrodynamics---plasmas}

\section{Introduction}
The non-luminous matter concentrated within the inner 0.015 pc at the Galactic Center 
has a measured mass of $2.6\pm0.2\times 10^6\;M_\odot$ (Genzel et al. 1996; Eckart 
\& Genzel 1996; Eckart \& Genzel 1997; Ghez et al.  1998).  Most of it 
appears to be associated with Sgr A*, a bright, compact radio source 
(Balick \& Brown 1974) that anchors the stars and gas locked in its vicinity,
and provides possibly the most compelling evidence for the existence of 
supermassive black holes. 

The spectrum of this unusual object can be described as a
power-law with an index $a$ that varies within the range $0.19-0.34$
($S_{\nu}\propto\nu^a$) from cm to mm wavelengths. However, one of 
the most interesting features currently under focus is the suggestion of a 
sub-millimeter (sub-mm) spectral bump (Zylka et al. 1992; Zylka et al. 
1995), since the highest frequencies appear to correspond to the 
smallest spatial scales (e.g., Melia, Jokipii \& Narayanan 1992; Melia 1994;
Coker and Melia 2000). In the case of Sgr A* one expects the sub-mm
emission to come directly from the vicinity of the black hole.
The existence of this bump (or ``excess'')
has been uncertain due to the variability of Sgr A*, but is now well established 
following a set of simultaneous observations (from $\lambda$20cm to $\lambda$1mm) 
using the VLA, BIMA, Nobeyama 45 m, \& IRAM 30 m telescopes (Falcke, et al. 1998).

The behavior of Sgr A* is dictated by the manner with which 
plasma accretes onto it from the nearby environment.
Three-dimensional hydrodynamic simulations of the gas dynamics at the location of
Sgr A* (Coker \& Melia 1997) indicate that the accreted specific angular momentum 
$\lambda$ (in units of $cr_S$, where $r_S\equiv 2GM/c^2$ is the Schwarzschild radius 
in terms of the black hole mass $M$) can vary by 
$50\%$ over $\simlt$ 200 years with an average equilibrium value for $\lambda$ of 
$40 \pm 10$. Thus, even with a possibly large amount of angular momentum 
present in the wind surrounding the nucleus, relatively little specific angular 
momentum is accreted.  This is understandable since clumps of gas with a high specific 
angular momentum do not penetrate to within the capture radius, $R_A\equiv 
2GM/{v_w}^2$, defined in terms of the wind velocity $v_w$ at infinity.  The 
variability in the sign of the components of $\lambda$ suggests that if an 
accretion disk forms at all, it dissolves, and reforms (perhaps) with an 
opposite sense of spin on a time scale of $\sim 100$ years.

The captured gas is highly ionized and magnetized, so it radiates via
brems\-strahlung, cyclo-synchrotron and inverse Compton processes.
However, this emissivity appears to be inefficient in the case of
Sgr A*, so most of the dissipated energy within the large scale
quasi-spherical inflow is carried inwards (Shapiro 1973;  Ipser \& Price 1977;
Melia 1992).  A viable explanation for Sgr A*'s low radiating efficiency 
is that the advected magnetic field is well below its equipartition value.
This may not be surprising in view of the 
fact that the actual value of $B$ depends strongly on the mechanism of field 
line annihilation, which is poorly understood.  Two processes that have been
proposed are (i) the Petschek (1964) mechanism, in which dissipation of the
sheared magnetic field occurs in the form of shock waves surrounding special
neutral points in the current sheets and thus, nearly all the dissipated 
magnetic energy is converted into the magnetic energy carried by the emergent 
shocks; and (ii) van Hoven's (1979) tearing mode instability, which relies 
on resistive diffusion of the magnetic field and is very sensitive to the 
physical state of the gas.  In either case, the magnetic field dissipation rate
is a strong function of the gas temperature and density, so that assuming
a fixed ratio of the magnetic field to its equipartition value may not
be appropriate.

Kowalenko \& Melia (1999) have used the van Hoven prescription to calculate
the magnetic field annihilation rate in a cube of ionized gas being compressed
at a rate commensurate with that expected for free-fall velocity onto the
nucleus at the Galactic Center.  
Whereas the rate of increase $\partial B/\partial t|_f$ in
$B$ due to flux conservation depends only on the rate $\dot r$ of the gas,
the dissipation rate $\partial B/\partial t|_d$ is a function
of the state variables and it is therefore not necessarily correlated with
$\dot r$.  Although these attempts at developing a physical model for
magnetic field dissipation in converging flows is still rather simplistic,
it is apparent from the test simulations that the equipartition assumption is not
always a good approximation to the actual state of a magnetohydrodynamic flow,
and very importantly, that the violation of equipartition can vary in degree
from large to small radii, in either direction.  

The first serious attempt at modeling the spectrum of Sgr A* using a
sub-equipartition profile for the magnetic field was made by
Coker \& Melia (2000), who adopted a spherical infall as
a simplified version of the actual accretion picture. Of course, the real
accretion flow will deviate from radial at small distances from the black
hole, where the gas begins to circularize with its advected specific angular
momentum. This deviation of the accreting gas away from a purely radial
infall may provide a clue for the appearance of the bump in the spectrum at 
sub-mm wavelengths, which seems to hint at the existence of a distinct geometry for
the emitter at this energy. We here suggest that the 
sub-mm ``excess'' in the spectrum of Sgr A* may be the first indirect evidence 
for the anticipated circularization of the gas falling into the black hole at 
$5-25\;r_S$. Although the physical conditions within the quasi-spherical
infall evidently suppress the magnetic field well below its equipartition
value, this need not be the case once the gas circularizes and forms a Keplerian
structure.  Indeed, it is expected that a magnetic dynamo within the differentially
rotating region may overwhelm the field annihilation rate and actually lead to
a saturated field intensity.

It is our intention here to fully explore the magnetic properties of this
inner circulating region, and to assess the likelihood of producing the
sub-mm spectral bump with a magnetic dynamo. Given the wide latitude of possible
configurations in the outer quasi-spherical infall for producing the longer 
wavelength radio emission, we will not here attempt to construct a complete
picture for the whole accretion region.  Rather, this task is better coupled  
to a hydrodynamic simulation that self-consistently merges the large scale
flow to the circularized structure at smaller radii.  These calculations
are now in progress, and their results will be reported in the future.

We will first summarize the key physical principles underlying the dynamo
process, and then examine what configuration the magnetic field should
have in the Keplerian region surrounding Sgr A*.  We will then demonstrate
that an excellent fit to the sub-mm data is possible with this picture,
and discuss the implications of this model to our overall understanding of
the environment surrounding the massive black hole at the Galactic Center. 

\section{The Magnetohydrodynamic Dynamo in a Rotational System}
Since the discovery of the local shear instability in weakly magnetized disks (Balbus \&
Hawley 1991, hereafter BH1), several numerical simulations (Hawley \& 
Balbus 1991, hereafter HB1; Hawley \& Balbus 1992, hereafter HB2; Hawley, Gammie \&
Balbus 1995, hereafter HGB1; Stone, Hawley, Gammie \& Balbus 1996, hereafter SHGB) 
have confirmed the fact that this instability plays a crucial role in rotational 
accretion systems. A linear analysis (Balbus \& Hawley 1992, hereafter BH2) has
demonstrated that the instability is extremely powerful. Its maximal growth rate is 
of the order of the angular velocity, provided that the latter decreases outward and 
that initially there exists a weak magnetic field.  In the case of an axisymmetric 
perturbation with a weak $B_z$ component of the magnetic field (BH1), the maximal growth rate in
a Keplerian disk reaches $0.75\,\Omega$ at $k_z\simeq{\Omega/v_{Az}}$, where
$\Omega$ is the angular velocity, $k_z$ is the $z$-component of the perturbation 
wavenumber and $v_{Az}=\sqrt{B^2_z/4\pi\rho}$ is the Alfv\'{e}n speed in the $z$-direction.
The numerical simulations 
show that the instability saturates at a turbulent state, producing a significant angular 
momentum flux, which is dominated by the Maxwell stress rather than the Reynolds stress. 
This process has been invoked to account for the origin of the anomalous viscosity in these
systems (SHGB).

Another important consequence of this instability, one that we explore at length in
this paper, is the amplification of the magnetic field. It has been shown that this 
instability constitutes a magnetohydrodynamic dynamo (Hawley, Gammie \& Balbus 1996, 
HGB2;  Brandenburg, Nordlund, Stein \& Torkelsson 1995, hereafter BNST). An external magnetic field, 
which is sometimes invoked to magnetize the disk, is not necessary for the instability to work. 
An internal turbulent  magnetic field can also drive the instability. Once one introduces 
the complete magnetohydrodynamic equations to describe a shearing hydromagnetic structure, 
the system saturates at a turbulent state that includes a significant magnetic field
energy density.  Numerical difficulties with the simulations have thus far prevented the 
acquisition of quantitative results that may be used directly to describe a real 
astrophysical situation.  However, based on what is now understood about the dynamo
process, some very important qualitative results have been obtained, and these can
provide a guide to the manner in which the magnetic field grows within the converging
flow around a compact object. 

\subsection{The Basic Equations}
For the sake of completeness, let us first analyze how the instability develops in a
weakly magnetized Keplerian flow. The basic dynamical equations are (BH1):
\begin{eqnarray}
{d\ln\rho\over dt}+{\bf \vec\nabla\cdot v}&=&0, \label{b1}\\
{d{\bf v}\over dt}+{\bf \vec\nabla\Phi}
&=&{1 \over 4\pi\rho}({\bf B\cdot\vec\nabla}){\bf B}-{1\over \rho}{\bf \vec\nabla}\left(P+{B^2\over
8\pi}\right), \label{b2}\\
{\partial{\bf B}\over \partial t}&=&({\bf B\cdot\vec\nabla}){\bf v}-({\bf \vec\nabla\cdot v}
){\bf B}-({\bf v\cdot\vec\nabla}){\bf B}\;,\label{b3}
\end{eqnarray}
where ${d/dt}$ is the Lagrangian derivative and ${\bf \Phi}$ is the external
gravitational potential. The other symbols have their usual meaning. We will adopt 
standard cylindrical coordinates $(r,\ \phi,\ z)$, where $r$ is the perpendicular 
distance from the $z$-axis.

Because the maximal growth rate is reached in the axisymmetric case with a weak 
vertical field, i.e., with ${\bf B}=(0,0,B_z)$, we will here consider 
perturbations for this specific situation.  As such, the Eulerian perturbations, 
which we denote by $\delta v$, $\delta B$ etc., are modulated by the function 
$e^{i(k_r r+k_z z-\omega t)}$, where $k_r$ and $k_z$ are, respectively, the radial 
and vertical components of the wavevector. The numerical simulations show that 
buoyancy is not a significant factor influencing the instability, nor is the
compressibility of the fluid. By neglecting these terms and assuming incompressibility,
so that $\delta \rho=0$ in all the equations other than the equation of motion and the 
equation of state (Balbus \& Hawley 1991), and that $\delta P=0$ in the equation 
of state, one obtains the following linearized dynamical equations:
\begin{eqnarray}
k_r\,\delta v_r+k_z\,\delta v_z &=& 0, \label{l1}\\ 
{\partial\,\delta v_r\over \partial t}-2\Omega\,\delta v_\phi&=&i{k_zB_z \over 4\pi\rho}\delta
B_r-ik_r\left({\delta P\over \rho}+{B_z\,\delta B_z\over 4\pi\rho}\right), \label{l2}\\
{\partial\,\delta v_z\over \partial t}&=&-ik_z{\delta P
\over \rho}, \label{l3}\\
{\partial\,\delta v_\phi\over \partial t}+{\kappa^2\over 2\Omega} \delta v_r&=&i{k_zB_
z \over
4\pi\rho}\delta
B_\phi, \label{l4}\\
{\partial\,\delta B_r\over \partial t}&=&ik_zB_z\,\delta v_r, \label{l5}\\
{\partial\,\delta B_z\over \partial t}&=&ik_zB_z\,\delta v_z, \label{l6}\\
{\partial\,\delta B_\phi\over \partial t}&=&{r\,d\Omega\over dr}\delta B_r+ik_zB_z\,\delta v_\phi\;, \label{l7}
\end{eqnarray}
where $\Omega$ is the angular velocity in the circularized flow, and
$\kappa^2=(2\Omega/r)\;d(r^2\Omega)/dr$ is the square of the epicyclic frequency.

Replacing the Lagrangian derivatives with respect to time $t$ by $-i\omega$ in the linearized
equations and eliminating the Eulerian perturbations, one obtains the dispersion relation
\begin{equation}
(\omega^2-k_z^2\,v_{Az}^2)^2-{k_z^2\over
k^2}\,\kappa^2(\omega^2-k_z^2\,v_{Az}^2)-4\Omega^2\,{k_z^4\,v_{Az}^2\over k^2}=0,
\end{equation}
where $k^2=k_z^2+k_r^2$.
In the case of Keplerian flows, $\kappa=\Omega$. This equation can easily be solved
for $\omega$, which yields
\begin{equation}
\omega^2_0=k_{z0}^2+{k_{z0}^2\over 2\,k^2_0}-2\sqrt{{k_{z0}^4\over k^2_0}+{k_{z0}^4\over 
16\,k^4_0}}\;,
\end{equation}
where $\omega_0\equiv\omega/\Omega$ and $k_0$ and $k_{z0}$ are, respectively,
$k$ and $k_z$ expressed in units of
$\Omega/v_{Az}$.  It is noted that $\omega^2$ reaches its minimum value of
$-9\Omega^2/16$ when $k^2=k_z^2=(15/16)(\Omega/v_{Az})^2$. For $k_r=0$, the modes become
stable when $k_z^2 > 3\,(\Omega/v_{Az})^2$, and in the long wavelength limit $\omega^2\simeq
-3\,(v_{Az}\,k_z)^2$.

In order to appreciate the physical implication of this instability, we examine the fastest
growing mode, which occurs when $k^2=k_z^2=(15/16)(\Omega/v_{Az})^2$ and $\omega^2=-9\Omega^2/16$.
Solving the linearized equations (\ref{l1})--(\ref{l7}), we get 
\begin{eqnarray}
\delta v_r&=&\delta v_\phi, \label{s1} \\
\delta B_r&=&-\delta B_\phi, \label{s2} \\
\delta B_r&=& i{4\over 3}{k_zB_z \over \Omega} \delta v_r, \label{s3} \\
\delta B_\phi&=& -i{5\over 4}{4\pi\rho\Omega\over k_zB_z} \delta v_\phi, \label{s4} 
\\
{|\delta B_r|^2\over 8\pi}&=&{5\over 3}{\rho |\delta v_r|^2\over 2}\;. \label{s5}
\end{eqnarray}
Keeping the time derivative terms in the linearized equations (\ref{l1})--(\ref{l7}) and
using this solution, we interpret the physics of this unstable mode as follows:
The perturbation $\delta v_r$, which is generated by $\delta v_\phi$ through the 
Coriolis force term $2\,\Omega\,\delta v_\phi$ in Equation (\ref{l2}), induces 
the perturbation $\delta B_r$ through Equation (\ref{l5}). Due to the shearing of 
the disk, the term $r\,d\Omega/dr\,\delta B_r$ in
Equation (\ref{l7}) then shows that $\delta B_r$ leads to the
production of a perturbation $\delta B_\phi$, which in turn enhances 
$\delta v_\phi$ through the right hand side of Equation (\ref{l4}). 
Thus a positive feedback loop is established.  Some of the other terms in the 
linearized equations act to stabilize the perturbation, but they are overwhelmed
by the positive feedback in the unstable modes. However, for the modes with a large 
wavenumber, the term $ik_zB_z\delta v_\phi$ in Equation (\ref{l7}) and the term 
$ik_zB_z \delta B_r/4\pi\rho$ in Equation (\ref{l2}) will overwhelm the positive 
feedback and make the mode stable.

\subsection{The Dynamo in Sgr A*}
The solution represented by Equation (\ref{s5}) deserves special attention;
it says that the turbulent kinetic energy density is approximately equal to the turbulent magnetic field 
energy density.  That is, it points to an equipartition  of kinetic and magnetic
field energy densities in the final saturated state of the system.  Several
numerical simulations (BNST, HGB1, HGB2, SHGB) have largely confirmed this
result.  They show that the ratio of these energy densities in the final 
turbulent state of the system is only weakly dependent on the initial and subsequent
physical conditions assumed in the calculations. We shall therefore here adopt the
conclusion drawn from these studies, that in rotational flows 
\begin{equation}
{<\delta B^2>\over 8\pi}=C_0\,{1\over 2}<\rho \delta v^2>\;,
\label{eq1}
\end{equation}
where the constant $C_0$ has a value between $1$ and $10$, depending on the vertical 
profile of the Keplerian structure. 

Although the above analysis is based on a very specific model, many numerical 
simulations (BNST, HGB2) have demonstrated that this instability exists more generally, 
and that even without an external magnetic field, it can produce a significant magnetic 
energy density, which therefore constitutes a hydromagnetic dynamo. 
We are here primarily interested in the final saturated state, whose existence
has been shown to be inevitable by the numerical simulations. In the following, we will build a
simple picture for the dynamo process in Sgr A*, and show how the turbulence is
sustained.  From these simulations, we infer that the vertical component of the internal 
turbulent magnetic field can produce unstable wave modes similar to those produced 
by an external magnetic field and that the dynamo is driven by them. Since we adopt a 
zero value for the external magnetic field (a reasonable approximation consistent
with the view that the turbulent magnetic field exterior to the Keplerian
flow is greatly sub-equipartition, Kowalenko \& Melia 1999), 
the turbulent magnetic field is in fact
then the total field ${\bf B}$.  Applying the above analysis to the situation in
Sgr A*, we replace the external vertical field with the mean-square root 
($\sqrt{<\delta B_z^2>} \equiv \sqrt{<B_z^2>}$)
of the turbulent magnetic field's vertical component (hereafter the symbol $<>$ denotes 
the mean value of the particular physical quantity).  However, because the final state is 
turbulent, it is expected that the growth rate will be smaller than that in the case
where there exists an underlying ordered external field. 

Before proceeding with the model, we need to examine whether or not this instability 
can be damped by Ohmic diffusion for the conditions expected in Sgr A*.  
Jin (1996) first discussed the possible damping 
of the shear instability using linear analysis.  More recently, this issue was
addressed numerically by Fleming, Stone and Hawley (2000). 
The overall conclusion is that the instability is effectively damped
when the diffusion length becomes comparable to the wavelength of the most 
unstable mode within the latter's growth time scale, i.e., roughly one
revolution period.  As we have seen, the most unstable wave mode in the Keplerian
flow is $k^2=k_z^2=15\Omega/16v_{Az}$, where $v_{Az}=\sqrt{<B_z^2>/4\pi <\rho>}$ 
is the vertical Alfv\'{e}n speed of the turbulent magnetic field.
As the magnetic field is relatively weak, $v_{Az}$ is small, and all the modes 
with $k_z<3\Omega/v_{Az}$ are unstable.  These unstable modes lead to an increase
in $<B^2_z>$, which in turn leads to an increase in the value of $v_{Az}$. 
So the wavelength $\lambda_z=2\pi/k_z$ of the most unstable mode increases.
There does not appear to be any physical reason why the magnetic field should
stop increasing before $\lambda_z$ has reached a value equal to the height $H$ of the 
Keplerian flow.  So the length scale of the most unstable wave mode is expected to be 
$H$.

Let us now show that the diffusion length scale within one period in Sgr A* is much 
smaller than this wavelength.  Using the parameter values characterizing the accretion onto 
this object, we find that the angular velocity of a Keplerian flow
near ten Schwarzschild radii ($r_S\equiv 2GM/c^2$, where for Sgr A*, $M\approx
2.6\times 10^6\;M_\odot$) is $\Omega\approx\sqrt{GM/ (10r_S)^3}\simeq 0.001$
s$^{-1}$.  
The temperature of the plasma in this region is expected to be around a few $10^{10}$ K 
(Melia 1994; Coker \& Melia 2000), so this gas is fully ionized, and its resistivity is (Lyman 1961)
\begin{equation}
\eta=7.26\times 10^{-9} {\ln\Lambda\over T^{3\over 2}}\qquad (\hbox{c.g.s.})\;, 
\end{equation}
where the Coulomb logarithm $\ln\Lambda$ is a slowly varying function of the
electron density and temperature; $30$ is a reasonable value for the conditions
at the Galactic Center.  Thus, for these parameter values, $\eta\simeq 0.22\times 
10^{-21}$ (c.g.s.).  The diffusion length scale corresponding to this resistivity
is 
\begin{equation}
L=\sqrt{\eta\, c^2\,\tau\over 4\pi}\;,
\end{equation}
where $\tau$ is the diffusion time, which we set equal to the growth time
scale of the most unstable wave mode, i.e., about $1000$ s here. Thus, we
infer a diffusion length of about $4$ cm, which is clearly much smaller than any
reasonable value of $H$, and therefore the characteristic wavelengths of the 
unstable modes.  We conclude that the dynamo in Sgr A* cannot be damped by 
Ohmic diffusion.

One of the important properties of the final turbulent magnetic field is that it
is dominated by its azimuthal component. All the simulations show that the azimuthal 
component counts for about 80 percent of the total magnetic field energy. 
This result can be understood by examining the equations describing the evolution
of the magnetic energy density (remembering still that the turbulent field generated
by the dynamo constitutes the total field in this system): 
\begin{eqnarray}
{1\over 2}{\partial B_\phi^2\over \partial t}&=& r{d \Omega\over dr}B_\phi B_r+B_\phi[{\bf
\vec\nabla\times(\delta v\times B)}]_\phi+{\eta c^2\over 
4\pi}B_\phi|\nabla^2{\bf B}|_\phi\;, \label{e1} \\
{1\over 2}{\partial B_r^2\over \partial t}&=& B_r[{\bf \vec\nabla\times(\delta 
v\times B)}]_r+{\eta c^2\over 4\pi}B_r|\nabla^2{\bf B}|_r\;, \label{e2} \\
{1\over 2}{\partial B_z^2\over \partial t}&=& B_z[{\bf \vec\nabla\times(\delta v\times B)}]_z+{\eta 
c^2\over 4\pi}B_z|\nabla^2{\bf B}|_z \label{e3}\;,
\end{eqnarray}
where $\eta$ is again the resistivity of the plasma, and $c$ is the speed of light.  
A linear analysis shows that the amplitudes of the azimuthal and radial components
of the magnetic field are equal when the perturbation is small, but that the final 
turbulent state is affected by the nonlinear character of the magnetohydrodynamic
equations.  As the amplitudes of the perturbation increase, nonlinear effects
become more important.  Due to the shearing of the Keplerian flow, the average value 
of $rB_\phi B_rd \Omega/dr$ is positive. The energy Equations (\ref{e1})---(\ref{e3}) 
therefore show that $r B_\phi B_rd \Omega /dr$ contributes to a growing anisotropy
of the turbulent magnetic field, in the sense that more and more magnetic field 
energy is generated in the azimuthal direction.  For a Keplerian flow with
$rd\Omega/dr=-3\Omega/2$, the growth rate due to the shearing of this structure is
larger than that associated with any other dynamo process.  So the azimuthal 
component of the magnetic field dominates the final magnetic field energy density. 

It is interesting to note that this simple scenario may also be used to interpret 
the anisotropic spectrum of the turbulent magnetic field energy density observed
in the numerical simulations (HGB1). Because the radial magnetic field is stretched 
in the azimuthal direction, the magnetic energy carried by the modes with the
largest wavelengths should be associated mostly with their azimuthal components. 
However the spectrum for the vertical and radial components should be similar
to each other.

So the situation with regard to Sgr A* is the following: 
as the gas flows inwards and spirals into an approximately Keplerian structure
at a distance from the black hole corresponding to the circularization radius,
a linear instability first stretches the magnetic field lines carried by the gas
and produces a radial component of ${\bf B}$. The magnetic field generated during 
this step is approximately in equipartition with the turbulent kinetic energy and 
counts for a small fraction of its final intensity, but it nonetheless
provides the seed for the next step. Second, the shearing in the Keplerian flow 
stretches the radial magnetic field in the azimuthal direction and it increases
the magnetic field energy (as seen, e.g., in HB2). The energy comes from the 
rotational energy of the gas, and during this process a significant amount of
angular momentum can be transported outward as a result of the Maxwell stress 
$B_rB_\phi/4\pi$.  Finally, some of the magnetic field energy is converted into
kinetic energy, which is eventually dissipated into thermal energy as a result
of the viscosity, through the Lorentz force term ${\bf B\cdot[\vec\nabla\times(\delta
v\times B)]}$. In addition, the magnetic field energy may be
dissipated through Ohmic resistivity.  For example, with their assumed large 
numerical resistivity, BNST found that about half of the magnetic field energy is 
dissipated in this way.  However, in the case of the Galactic Center, the 
actual resistivity is rather small and $\eta c^2/ 4\pi r_S\simeq 10^{-13}$ cm 
s$^{-1}$ is insignificant compared to any velocity scale within the plasma.
The dissipation of the magnetic field in this fashion is therefore negligible.

So from the magnetic energy Equations (\ref{e1})---(\ref{e3}), we infer the
following proportionality based on dimensional analysis:
\begin{equation}
{\sqrt{<\rho\,\delta v^2>}\over\sqrt{<\rho>}\; H} \propto r{d\Omega\over
dr}\;, \label{eq2}
\end{equation}
where $H$ is the height of the flow and we have used 
$\sqrt{<\rho\, \delta v^2>}/\sqrt{<\rho>}$ to characterize the turbulent velocity.
Combining Equations (\ref{eq1}) and (\ref{eq2}), we therefore obtain
\begin{equation}
\sqrt{<B^2>}\propto \sqrt{<\rho>}\, H\, r{d\Omega\over dr}\;,
\label{eq3}
\end{equation}
which is the main result we have been seeking.  Note that for a Keplerian flow,
$\Omega=(GM/r^3)^{1/2}$, for which $rd\Omega/dr=-(3/2)\,\Omega$.

Some support for the validity of this relationship has been provided by the simulations
reported in BNST, which include an analysis of the magnetohydrodynamic dynamo at two 
different radii, one of which is five times smaller than the other. Since this 
Keplerian flow is characterized by the following equations,
\begin{eqnarray}
(H\Omega)^2&\propto& P/\rho\;,\\ 
\rho&\propto& r^{15/8}\;,\\  
H&\propto& r^{9/8}\;,
\end{eqnarray} 
the expression in (\ref{eq3}) suggests that $<B^2>\propto P$, the gas
pressure. 
An inspection of the numerical results in BNST shows that the ratio of magnetic 
field energy density to thermal energy density at the two radii is
0.013 and 0.014.  Given that the ratios of the densities, the heights and the
angular velocities at the two radii are, respectively, $25$, $5$, and
$10$, the constancy of the ratio of magnetic field energy density
to thermal energy density is a strong indication that Equation
(\ref{eq3}) provides an
adequate representation of the saturated magnetic field intensity under these
conditions.  The validity of this expression is also supported by the
global simulations of Hawley (2000). 

We conclude that the magnetic field intensity within the gas converging onto
Sgr A* approaches the functional form given in Equation (\ref{eq3}) once the
flow
settles into a Keplerian structure at small radii.  The fluctuations associated
with the accreted angular momentum (Coker \& Melia 1997) correspond to a circularization
radius $\sim 5-50\;r_S$, so we anticipate that ${\bf B}$ will approach
the distribution in Equation (\ref{eq3}) within this region. 

\section{Calculation of the Spectrum}
\subsection{The Anomalous Viscosity}
In a Keplerian flow with column density $\Sigma$ and angular velocity
$\Omega=(GM/r^3)^{1/2}$, the radial velocity $v_r$ at (cylindrical)
radius $r$ is given as (e.g., Stoeger 1980)
\begin{equation}
v_r=-{3\over r^{1/2}\Sigma}\,{\partial\over\partial r}\left(\nu\Sigma r^{1/2}\right)\;,\label{vr}
\end{equation}
where $\nu$ is the kinematic viscosity, 
\begin{equation}
\nu={2\over 3}{W_{r\phi}\over\Sigma\;\Omega}\;,
\end{equation}
and $W_{r\phi}$ is the vertically integrated sum of the Maxwell
and Reynolds stresses (Balbus et al. 1994).  For the problem at hand, 
the Maxwell stress dominates, and
\begin{equation}
W_{r\phi}\approx\beta_\nu\int dz\;\langle{B^2\over 8\pi}\rangle\label{visc}
\end{equation}
(the average inside the integral being taken over time).
Even though this approximation is valid simply on the basis that the Reynolds
stress is relatively small, its validity is enhanced by the fact that since the turbulent 
velocity (which accounts for this kinetic stress) and ${\bf B}$ are generated by the 
same process, both should be scalable by ${\bf B}$.  Numerical simulations (e.g., by 
Brandenburg, et al. 1995) show that $\beta_\nu$ changes very slowly with $r$.  In the 
particular cases considered by these authors, $\beta_\nu$ ranged in value from $\approx 0.1$ to
$0.2$, which represents an increase by a factor of only $2$ as $r$ decreased
by a factor of $5$.  We will here find that a ``mean'' value of $\sim 0.2-0.3$ is
required for this quantity.

To use the result of \S\ 2.2, we first need to know the vertical structure of the
Keplerian flow. For steady conditions, one can obtain the vertical profile by
assuming that the gas is (on average) in local hydrostatic equilibrium. Balancing 
gravity and the pressure gradient in the vertical direction,
we obtain the scale height
\begin{equation}
H=\sqrt{2 R_g T r^3\over \mu GM}\; ,
\label{H}
\end{equation}
where $T$ is the gas temperature at radius r, $R_g$ is the gas constant, 
and $\mu$ is the molecular weight.
As an approximation, we will assume that the Keplerian flow is axisymmetric and
is independent of the vertical coordinate. Written another way, we have
\begin{equation}
(H\Omega)^2=2P/\rho\; .
\end{equation}
Following the result of \S\ 2.2, we will write
\begin{equation}
\int dz\;\langle{B^2\over 8\pi}\rangle\approx
\beta_p\int P\;dz=\beta_p{R_g\Sigma T\over\mu}\;, \label{mag}
\end{equation}
where $\beta_p$ is roughly constant with a value of $\approx 0.03$.
In our simulations, we will find that a value $\sim 0.03-0.04$ is required
to match the observations.  Thus, with 
\begin{equation} 
\dot M=-2\pi\,r\,\Sigma\,v_r\;,\label{mdot}
\end{equation}
we can integrate Equation (\ref{vr}) to obtain
\begin{equation}
v_r={rT(r)\over T_0r_0/v_{r0}+
\mu(GM)^{1/2}(r_0^{1/2}-r^{1/2})/\beta_\nu\beta_p
R_g}\;, \label{vr1}
\end{equation}
where the quantities with subscript $0$ are to be
evaluated at the outer edge of the Keplerian flow
(i.e., at radius $r_0$).

\subsection{Energy Equations for the Keplerian Flow}

In order to get the radial dependence of the flow, we need to solve the 
energy equation to determine the temperature. The gas is heated primarily 
by the viscous dissipation of turbulent energy and Ohmic dissipation 
(which converts magnetic field energy directly to thermal energy), and is cooled by 
synchrotron and bremsstrahlung emission. 
Equation (\ref{b1}) is always valid. By incorporating Ohmic diffusion and
viscous dissipation, Equations (\ref{b2}) and (\ref{b3}) take the form:
\begin{eqnarray}
{d (\rho{\bf v})\over dt}+\rho{\bf \vec\nabla\Phi}+\rho{\bf v (\vec\nabla\cdot v)}
&=&{1 \over 4\pi}({\bf B\cdot\vec\nabla}){\bf B}-{\bf
\vec\nabla}\left(P_{th}+{B^2\over 8\pi}\right)+{2\bf \vec\nabla\cdot (S}\rho\nu)\; ,
\label{b22}\\
{\partial{\bf B}\over \partial t}&=&{\bf \vec\nabla \times (v\times B)}-{c^2\eta\over
4\pi}{\bf\vec\nabla\times(\vec\nabla\times B)}\; ,\label{b33}
\end{eqnarray}
respectively, where $S_{ij}=(1/2) 
[v_{i,j}+v_{j,i}-(2/3)\delta_{ij}v_{k,k}]$ and $P_{th}=\rho R_g
T/\mu+P_{rad}$ is the
non-magnetic pressure. For a Keplerian flow,
$S_{r\phi}=-(3/4)\Omega$. The radiation pressure, $P_{rad}$, is given
in the Rayleigh-Jeans
approximation:
\begin{equation}
P_{rad}={8\pi\over 9} kT\left({\nu_m\over c}\right)^3\; ,
\end{equation}
where $\nu_m$ is the frequency below which the radiative emission is highly absorbed, so
that the optical depth from $r$ to infinity is unity (see Melia 1994). 

Projecting Equation (\ref{b22}) onto the vector
${\bf v}$ and Equation (\ref{b33}) onto the vector ${\bf B}$, we eventually find that 
\begin{eqnarray}
{\partial\over \partial t}\left[\rho\left({1\over 2}v^2+\Phi\right)\right]&=&-{\bf
\vec\nabla\cdot}\left[\rho{\bf
v}\left({1\over
2}v^2+\Phi\right)\right]-{\bf v\cdot\vec\nabla}P_{th}+{2\bf v\vec\nabla\cdot
(S}\rho\nu)\nonumber \\
&&-{1\over 4\pi}{\bf v\cdot [B\times(\vec\nabla\times B)]}\; , \label{b222}\\ 
{\partial\over \partial t}\left({B^2\over 8\pi}\right)&=&{1\over 4\pi}{\bf
B\cdot[\vec\nabla\times(v\times
B)]}+{c^2\eta\over 16\pi^2}{\bf\vec\nabla\cdot[B\times(\vec\nabla\times B)]}-\eta {\bf
J}^2\; ,
\label{b333} 
\end{eqnarray}
where ${\bf J}=(c/4\pi){\bf \vec\nabla\times B}$ is the current density.
Adding the two equations gives
\begin{eqnarray}
{\partial\over \partial t}\left[{B^2\over 8\pi}+\rho\left({1\over
2}v^2+\Phi\right)\right]&=&-{\bf
\vec\nabla\cdot}\left[-2\nu \rho{\bf v\cdot S}+{1\over
4\pi}{\bf B\times}\left({\bf v\times B}-{c^2\eta\over 4\pi}{\bf \vec\nabla\times
B}\right) \right.  \nonumber \\
&&\left.+\rho{\bf v}\left({1\over
2}v^2+\Phi\right)\right] -\eta{\bf J}^2-2\nu \rho {\bf S}^2-{\bf v\cdot\vec
\nabla}P_{th}\; .\label{eng1}
\end{eqnarray}
In steady state, the heating term is therefore inferred to be 
\begin{eqnarray}
\Gamma&\equiv&\eta{\bf J}^2+2\nu \rho {\bf S}^2 \nonumber \\
      &=&-{\bf \nabla}\cdot\left[\rho{\bf v}\left({1\over 2}v^2+\Phi\right)+{1\over
4\pi}{\bf B\times(v\times B}-{c^2\eta\over 4\pi}{\bf \vec\nabla\times B)}-2\nu \rho{\bf v\cdot
S}\right]-{\bf v\cdot\vec \nabla}P_{th}\;. \label{h1}
\end{eqnarray}
Following the argument by Balbus  et al. (1994), we see that the divergence of the viscous flux and
Ohmic flux is insignificant, and so we can neglect these terms in the following discussion.

In a Keplerian flow, the velocity fluctuations are small compared to the azimuthal component of the
velocity and it is the correlated fluctuations in the velocity and magnetic field components that
produces the anomalous viscosity (Balbus et al. 1994). So ${\bf v}=v_\phi {\bf e_\phi}+\delta{\bf 
v}$, with $\delta v$ much smaller than $v_\phi\equiv\sqrt{GM/r}$.  Introducing this into
Equation (\ref{h1}) and neglecting the high order terms, we get the steady state heating
rate
\begin{eqnarray}
\Gamma&=&-{\bf v\cdot\vec\nabla}P_{th}-{1\over Hr}{\partial\over \partial
r}\left({v_rHr<B^2>\over 4\pi}\right)  \nonumber \\
& &-{1\over Hr} {\partial\over \partial r}\left[Hr\rho v_r\left({1\over 
2}{v_\phi}^2+{1\over 2}{v_r}^2+\Phi\right)+Hr v_\phi\left<\rho v_r\delta
v_\phi-{B_rB_\phi\over 4\pi}\right>\right] \label{h2}\; ,
\end{eqnarray}
where the correlated fluctuation on the right hand side is given by
\begin{equation}
\left<\rho v_r\delta v_\phi-{B_rB_\phi\over
4\pi}\right>={W_{r\phi}\over
2H}\; . \label{h3}
\end{equation}

Next, the temperature can be determined by solving the thermal energy conservation equation:
\begin{equation}
{\partial (\rho\epsilon)\over \partial
t}=-{\bf\vec\nabla\cdot(v}\rho\epsilon)+\Gamma-\Lambda-P_{th}{\bf \vec \nabla
\cdot v}\; , \label{eng2}
\end{equation}
where $\Lambda$ is the cooling rate, and $\rho\epsilon=\alpha nkT+3P_{rad}$ is the thermal and radiation
energy density.
In the fully ionized but non-relativistic limit, $\alpha=3$, whereas in the relativistic electron limit,
$\alpha=9/2$.
Adding Equations (\ref{eng1}) and (\ref{eng2}), we have the complete energy conservation
equation of the system
\begin{eqnarray}
{\partial\over \partial t}\left[{B^2\over 8\pi}+\rho\left({1\over
2}v^2+\Phi+\epsilon\right)\right]&=&-\eta{\bf J}^2-2\nu
\rho {\bf S}^2-\Lambda-{\bf
\vec\nabla\cdot}\left[\rho{\bf
v}\left({P_{th}\over \rho}+{1\over
2}v^2+\Phi+\epsilon\right)\right.\nonumber \\
&&\left.-2\nu \rho{\bf v\cdot  S}+{1\over
4\pi}{\bf B\times(v\times B}-{c^2\eta\over 4\pi}{\bf \vec\nabla\times B)} \right]+\Gamma\; . 
\label{eng}
\end{eqnarray}

In steady state, the derivative with respect to time equals zero. So using the mass conservation
Equation (\ref{mdot}) for a Keplerian flow and Equations (\ref{h2}), (\ref{h3}), (\ref{visc}) and
(\ref{mag}) , we have
\begin{eqnarray}
\rho v_r{\partial \epsilon\over \partial r}&=&\Gamma-\Lambda-P_{th}{\bf \vec \nabla
\cdot v}
\nonumber \\
&=&-\Lambda-{\bf \vec \nabla \cdot (v} P_{th})+{1\over Hr}{\partial\over
\partial r}\left({\beta_pR_gT\dot{M}\over 2\pi \mu}\right) \nonumber\\
&&+{1\over Hr}{\partial\over
\partial r}\left[{\dot{M}\over 4\pi }\left({1\over
2}({v_\phi}^2+v_r^2)+\Phi\right)+{\dot{M}\beta_\nu\beta_pR_gTv_\phi\over 4\pi 
v_r\mu}\right]\; .
\end{eqnarray}
The temperature equation follows from this once we substitute for
$H$, $\epsilon$, $P_{th}$ etc., whence 
\begin{equation}
C_1{T^\prime\over T}=-{\Lambda\over \rho v_r}+C_2\; ,
\label{tem}
\end{equation}
where 
\begin{eqnarray}
C_1&=&\epsilon+{P_{th}\over\rho}-{1\over
2}v_r^2+{2\beta_pR_gT\over\mu}+{3\beta_\nu\beta_pR_gTv_\phi\over 2\mu
v_r}\; , \nonumber \\
C_2&=&{5\over 2r}v_r^2-{GM\over
2r^2}+{\rho^\prime\over \rho}\left[{32\pi
R_gT\over 9n}\left({\nu_m\over
c}\right)^3+v_r^2-{\beta_\nu\beta_pR_gTv_\phi\over \mu
v_r}\right]\nonumber \\
&&-{32\pi R_gT\over 3n}\left({\nu_m\over
c}\right)^3{\nu_m^\prime\over\nu_m} -{\beta_\nu\beta_pR_gTv_\phi\over \mu  
v_r}{2\over r}\; .
\end{eqnarray}
The $^\prime$s denote derivatives with respect to $r$.

Now, from Equations (\ref{H}), (\ref{mdot}) and (\ref{vr1}), we find that
\begin{eqnarray}
{T^\prime\over 2T}+{3\over 2r}&=&{H^\prime\over H}\; ,\\
{1\over r}+{v_r^\prime\over v_r}+{H^\prime\over H}+{\rho^\prime\over \rho}&=&0\; ,\\
{T^\prime\over T}+{1\over r}+{\mu\sqrt{GM}v_r\over
2\beta_\nu\beta_pR_gr\sqrt{r}T}&=&{v_r^\prime\over v_r}\; .
\end{eqnarray}
Thus,
\begin{equation}
{\rho^\prime\over \rho}=-{7\over 2r}-{3T^\prime\over 2T}-{\mu\sqrt{GM}v_r\over
2\beta_\nu\beta_pR_gr\sqrt{r}T}\; . 
\end{equation}

Therefore eliminating $\rho^\prime$ in Equation
(\ref{tem}), we get a differential equation for the temperature:
\begin{equation}
E_1T^\prime=E_2-{\Lambda\over \rho v_r R_g}-{32\pi T\over
3n}\left({\nu_m\over c}\right)^3{\nu_m^\prime\over \nu_m}\; ,
\label{tem0}
\end{equation} 
where 
\begin{eqnarray}
E_1&=&\alpha+2+{32\pi\over
9n}\left({\nu_m\over c}\right)^3+{2\beta_p\over\mu}+{3\beta_\nu\beta_pv_\phi\over 2\mu
v_r}-{v_r^2\over 2R_gT}+{3\over 2}E_3\; , \nonumber \\
E_2&=&-{2T\over r}{\beta_\nu\beta_pv_\phi\over \mu v_r}+\left({5\over 2r}v_r^2-{GM\over
2r^2}\right){1\over R_g}-E_3T\left({7\over 2r}+{\mu\sqrt{GM}v_r\over  
2\beta_\nu\beta_pR_gr\sqrt{r}T}\right)\; ,\nonumber \\
E_3&=&{v_r^2\over R_gT}-{\beta_\nu\beta_pv_\phi\over \mu v_r}+{32\pi\over
9n}\left({\nu_m\over
c}\right)^3\; . 
\end{eqnarray}
Using the characteristic parameters for Sgr A*, we find that the radiation energy density 
and pressure are always negligible compared to those for the gas within the Keplerian flow. 
So, in Equation (\ref{tem0}) we will neglect those terms that depend on $\nu_m$.
Thus this equation provides the temperature profile throughout the inner region once the
outer boundary conditions are specified.

\subsection{Calculation of the Spectrum}

The flux density (at earth) produced by the Keplerian portion of
the flow is given by
\begin{equation}
F_{\nu_0}={1\over D^2}\int I_{\nu^\prime} \left(1-{r_S\over r}\right)^{3/2}\ dA\, ,
\end{equation}
where $D=8.5$ kpc is the distance to the Galactic Center,
$\nu_0$ is the observed frequency at infinity and $\nu^\prime$ is
the frequency measured by a stationary observer in the Schwarzschild frame.
(For simplicity, we here assume the metric for a non-spinning black hole.
A more thorough exploration of the parameter values, including the black
hole spin, will be discussed elsewhere.) The frequency transformations
are given by
\begin{equation}
\nu_0 = \nu^\prime \sqrt{1-r_S/r}\;,
\end{equation}
\begin{equation}
\nu^\prime = \nu {\sqrt{1-v_\phi^2/c^2}\over 1-(v_\phi/c)\cos{\theta}}\;,
\end{equation}
where $\nu$ is the frequency measured in the co-moving frame, and $\theta$ is the
angle between
the velocity $\vec v_\phi$ and the line of sight. Since the radial velocity is
always much smaller than $v_\phi$, we ignore this component in the transformation
equations. So $\cos{\theta} = \sin{i}\;\cos{\phi}$,
where $i$ is the inclination angle of the axis perpendicular to
the Keplerian flow, and $\phi$ is the azimuth of the emitting element.
When the Doppler shift is included, the blue shifted region is located primarily
near $\phi=0$ while the red shifted region is at $\phi=\pi$.
The other quantities that are necessary for an evaluation of the flux
density are the area element
\begin{equation}
dA = {1\over \sqrt{1-r_S/r}}\cos{i}\ r\ dr\ d\phi\;,
\end{equation}
and the specific intensity
\begin{equation}
I_{\nu^\prime} = B^\prime_{\nu^\prime}(1-e^{-\tau})\;, \label{Intensity1}
\end{equation}
where
\begin{equation}
B^\prime_{\nu^\prime}= \left({\sqrt{1-v_\phi^2/c^2}\over 
1-(v_\phi/c)\cos{\theta}}\right)^3 B_\nu\;,
\end{equation}
and the optical depth is
\begin{equation}
\tau=\int \kappa^\prime_{\nu^\prime}\;ds = \kappa_\nu\;{2H\over
\cos{i}}\;{1-(v_\phi/c)\cos{\theta}\over \sqrt{1-v_\phi^2/c^2}}\;, \label{depth}
\end{equation}
where $\kappa_\nu$ is the absorption coefficient.  In the case where
the optical depth $\tau\ll 1$, Kirchoff's law allows us to write
\begin{equation}
I_{\nu^\prime}\approx B^\prime_{\nu^\prime} \tau=\epsilon_\nu{2H\over
\cos{i}}\left({\sqrt{1-v_\phi^2/c^2}\over
1-(v_\phi/c)\cos{\theta}}\right)^2\;, \label{Intensity2}
\end{equation}  
where $\epsilon_\nu = B_\nu\ \kappa_\nu$ is the emissivity.
The presence of a substantial azimuthal component of the magnetic field
makes it convenient to calculate the observed flux directly from the
Extraordinary and Ordinary components of the intensity.  The most
convenient approach is to select the symmetry axis of the Keplerian
flow as the reference direction. The observed flux densities in the
azimuthal and the reference directions are given by
\begin{eqnarray}
F_{1\nu_0}&=& {1\over D^2}\int (I^e_{\nu^\prime}\cos^2{\phi^\prime}+
I^o_{\nu^\prime}\sin^2{\phi^\prime})\left(1-{r_S\over r}\right)^{3/2}\ dA\ ,\\
F_{2\nu_0}&=& {1\over D^2}\int (I^e_{\nu^\prime}\sin^2{\phi^\prime}+
I^o_{\nu^\prime}\cos^2{\phi^\prime})\left(1-{r_S\over r}\right)^{3/2}\ dA\ ,
\end{eqnarray}
respectively, where $\phi^\prime+\pi/2$ is the position angle of the
magnetic field vector within the emitting element that
has an azimuth of $\phi$, so that
$\cot{\phi^\prime}=\cot{\phi}\;\cos{i}$. $I^e_{\nu^\prime}$ and $I^o_{\nu^\prime}$ are
the
specific intensities for
the Extraordinary and Ordinary waves, respectively. For thermal synchrotron
radiation, the emissivities are given by (Pacholczyk 1970)
\begin{eqnarray}
\epsilon^e&=& {\sqrt{3} e^3\over 8\pi m_e c^2} B \sin{\theta^\prime} \int_0^\infty
N(E)[F(x)+G(x)]\ dE\ , \label{com1} \\
\epsilon^o&=& {\sqrt{3} e^3\over 8\pi m_e c^2} B \sin{\theta^\prime} \int_0^\infty
N(E)[F(x)-G(x)]\ dE\;, \label{com2}
\end{eqnarray}  
where $N(E)$ is the electron distribution function at energy $E$, and
\begin{eqnarray}
\cos{\theta^\prime}&=& {\cos{\theta}-v_\phi/c\over 1-(v_\phi/c)\cos{\theta}}\; ,\\
x&=&{4\pi\nu m_e^3c^5\over 3eB\sin{\theta^\prime}E^2}\; ,\\
F(x)&=& x\int_x^\infty K_{5/3}(z)\ dz\;, \\
G(x)&=& x\ K_{2/3}(x)\;.
\end{eqnarray}
$K_{5/3}$ and $K_{2/3}$ are the corresponding modified Bessel functions.
The total flux density produced by the Keplerian portion of the flow is
the sum of these two.

Given the temperature profile indicated in Figure 2, a non-negligible fraction 
of the radiation with frequency higher than $\sim 3\times 10^{11}$ Hz is self-Comptonized
by the hot plasma into the X-ray band.  We note that for the conditions
prevalent in this region (notably the temperature of the electrons and the
characteristic energy of the seed photons), the electron-photon scattering occurs
in the Thomson limit. Since in addition the electron thermal energy ($\sim kT$) 
greatly exceeds the kinetic energy associated with the radial and Keplerian 
motion of the gas, the electrons can be treated as locally isotropic.  In turn, 
the upscattered radiation will also be emitted isotropically.  Without loss of 
generality, we can therefore calculate
the photon number density function $n_\epsilon$ from the angle-integrated intensity
(where $n_\epsilon d\epsilon$ represents the number density of photons with
energies between $\epsilon$ and $\epsilon + d\epsilon$ and depends primarily on
the radius $r$).  Under these conditions, a single electron with Lorentz factor
$\gamma$ emits photons with energies between $\epsilon_s$ and
$\epsilon_s + d\epsilon_s$ at a rate  (e.g., Melia \& Fatuzzo 1989)
\begin{equation}
{dN\over dt d\epsilon_s} \left(\epsilon_s, \gamma\right)
= {3c\sigma_T\over 16}\;{\epsilon_s\over \beta\gamma^2}\;
\int_{-1}^1 d\mu_s \; \int_{\epsilon_l}^{\epsilon_u} d\epsilon
{n_\epsilon\over \epsilon^2} g(\mu', \mu_s')\;,
\end{equation}
where $\mu = \cos(\theta)$ and $\mu_s = \cos(\theta_s)$ represent
the lab-frame (unprimed) propagation angles (with respect to
the electron's direction of motion) of the incident and scattered
photons, respectively.  Here, we have made use of the transformations
$\epsilon_u = \epsilon_s \gamma^2 (1-\beta\mu_s)(1+\beta)$, and
$\epsilon_l = \epsilon_s {(1-\beta\mu_s)/(1+\beta)}$.
The function $g(\mu', \mu_s')$ takes into account the angular dependence
of the cross-section, and is given in terms of the rest-frame (primed) angles
by the expression
$g(\mu', \mu_s') = \left(1-\mu^{\prime 2}\right)\left(1-\mu_s^{\prime 2}\right)+(1/2)
\left(1+\mu^{\prime 2}\right)\left(1+\mu_s^{\prime 2}\right)$.
The rest frame and lab frame angles are related through the transformation
$\mu' = (\mu-\beta)/(1-\beta\mu)$.   Conservation of energy requires that
$\mu = \left\{1-(\epsilon_s/\epsilon)\left[1-\beta\mu_s\right]\right\}/\beta$.

With the upscattered photons emitted isotropically,
the total photon production rate is then found by integrating
over all of the electrons within the emission region, yielding
\begin{equation}
{dN_{TOT}\over dt d\epsilon_s} = \int_{gas} dV \int_0^\infty
dE \; n_e\;  f(E)\;\left[{dN \over dt d\epsilon_s}\right]\;,
\end{equation}
where $n_e$ is the (radially dependent) number density of
electrons and $f(E)$ is the relativistic Maxwell-Boltzmann (MB)
distribution function normalized to unity.  Note that the
temperature (and therefore the MB distribution function) is
also radially dependent.  

\subsection{Results}

Not all the solutions to Equation (\ref{tem0}) are physically acceptable. Equation
(\ref{vr1}) shows that for given parameters $\beta_p$ and $\beta_\nu$, some boundary values 
of $T_0$, $r_0$ and $v_{r0}$ will make the denominator vanish at a critical radius $r_{crit}$.
For a physically meaningful flow, we therefore need $r_{crit} \le r_i$, where $r_i$ is
the inner boundary of the Keplerian flow ( $r_i = r_S$ in the
following discussion ). 
For simplicity, we shall set $r_{crit}=r_i$, for which
\begin{equation}
{T_0 r_0\over v_{r0}}+{\mu(GM)^{1/2}(r_0^{1/2}-r_i^{1/2})\over \beta_\nu\beta_pR_g}= 
0\; .
\label{vr2}
\end{equation}
This fixes $T_0$ in terms of $r_0$ and $v_{r0}$.

A second constraint is provided directly by the current IR and UV observations of
Sgr A*, which rule out the possible presence of an optically thick disk.  
%(We shall see how this comparison works in greater detail below.) 
Working within these constraints, the best fit model for the sub-millimeter bump is that shown
in Figures 1-7. The latest Chandra observation (indicating a luminosity for Sgr A*
of no more than $2.66\times 10^{33}$ ergs s$^{-1}$ between $2$ and
$10$ keV) restricts the possible range of $\dot M$ near the black hole.
The reason for this can be understood qualitatively as
follows.

\begin{figure}[p]
  \centering
{\begin{turn}{-0}
  \psfig{figure=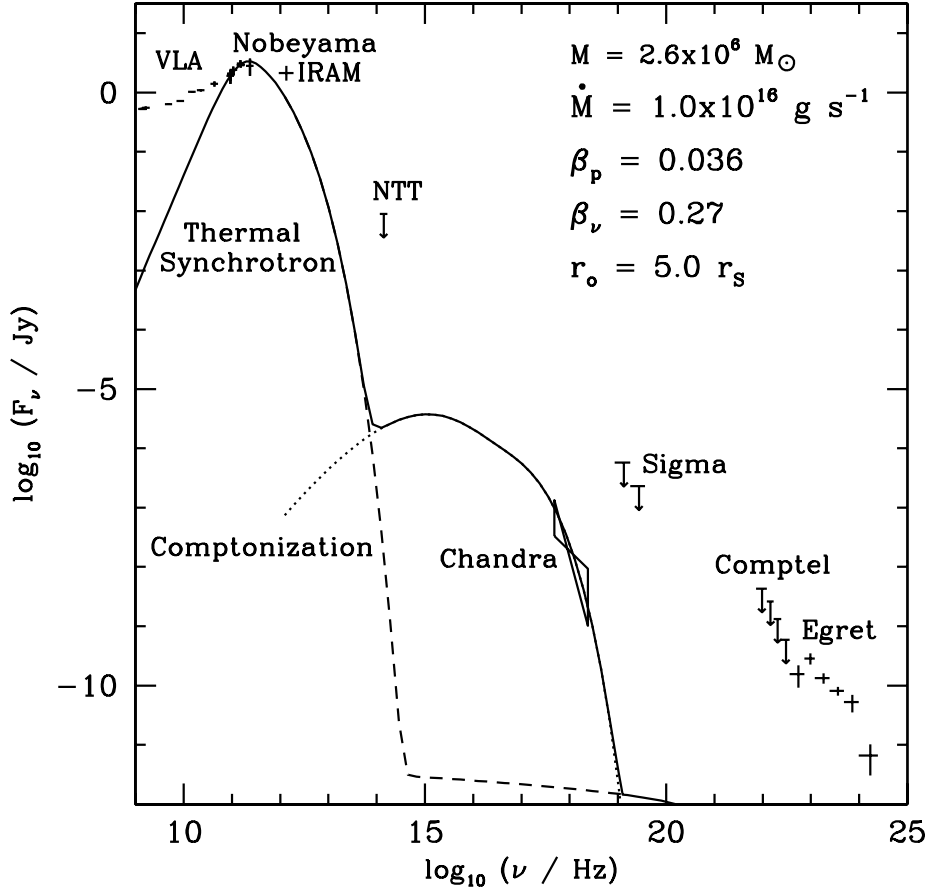,width=5.in}
\end{turn}}
  \caption{The spectrum corresponding to the best fit model, whose parameter values 
are indicated within the figure itself. The dashed curve corresponds to the thermal
synchrotron plus bremsstrahlung component, whereas the dotted curve shows the
self-Comptonized spectrum. The solid curve is the sum of these. 
The disk has an inclination angle of $60^o$. It is also necessary to 
specify the ratio of $v_r$ to its free-fall value at $r_0$.  For this model, this 
ratio is $5.0\times 10^{-4}$. The references for the data are given in the text.}
  \label{fig:specbest}
\end{figure}

\begin{figure}[p]
  \centering
{\begin{turn}{-0}
  \psfig{figure=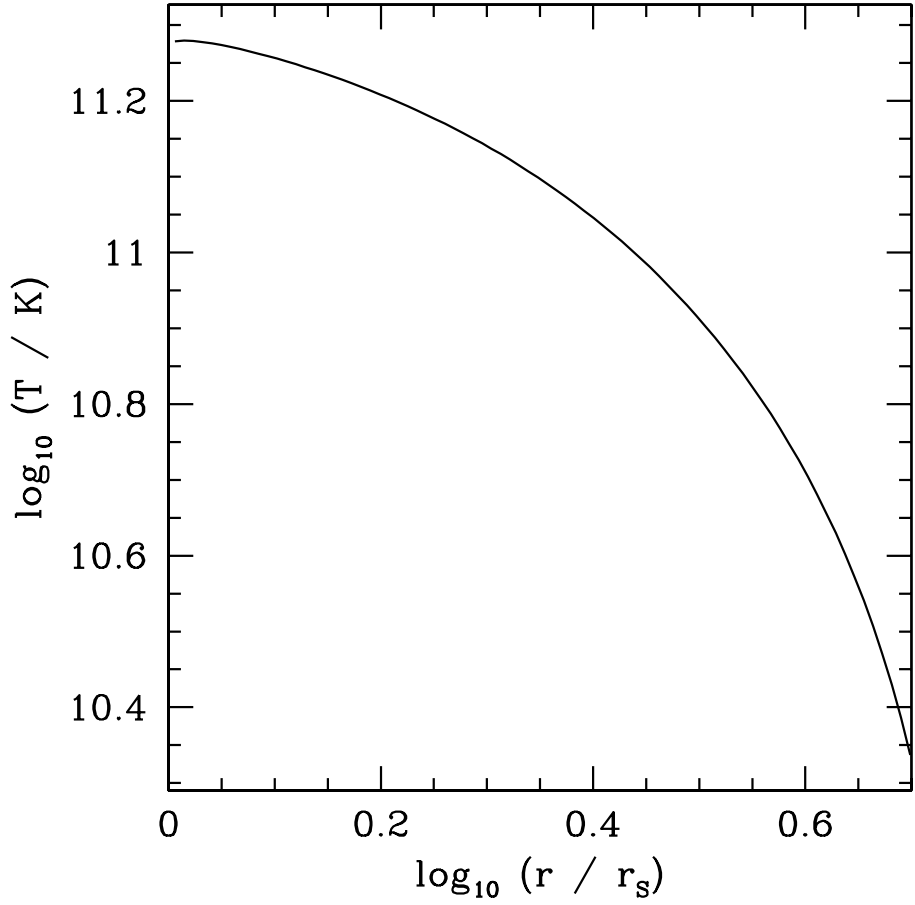,width=5.in}
\end{turn}}
  \caption{Temperature profile corresponding to the best fit model, shown in Figure 1.}
  \label{fig:specbestT}
\end{figure}

\begin{figure}[p]
  \centering
{\begin{turn}{-0}
  \psfig{figure=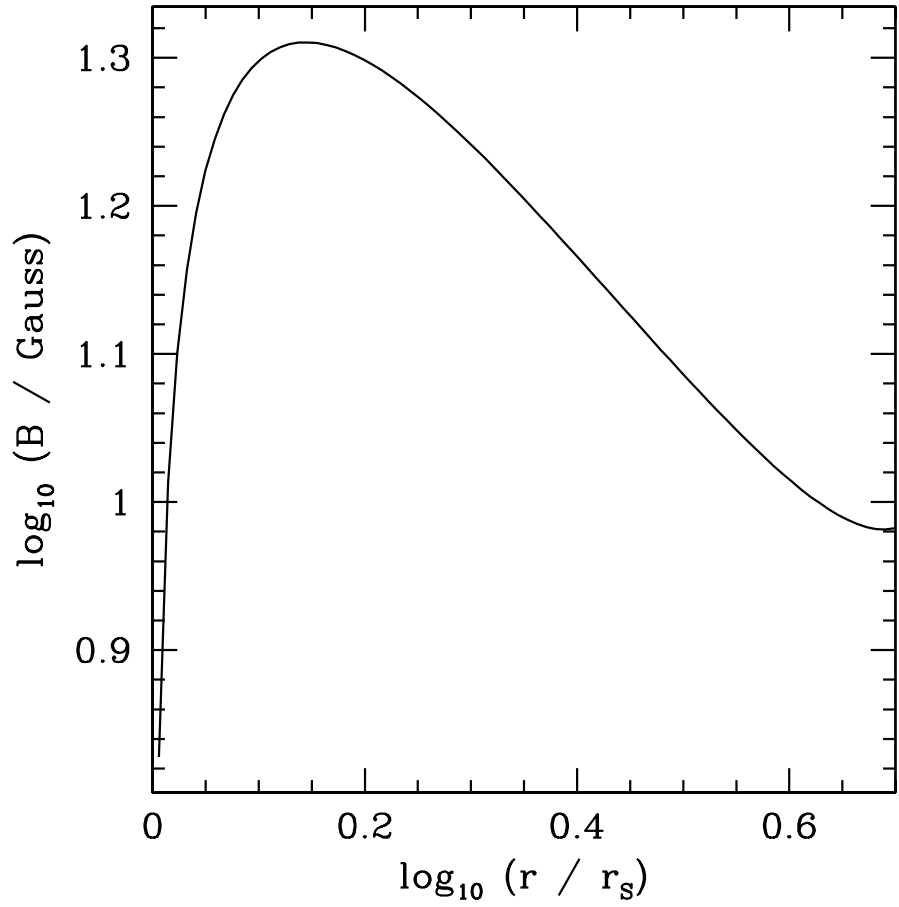,width=5.in}
\end{turn}}
  \caption{Magnetic field intensity corresponding to the best fit model, shown in Figure 1.}
  \label{fig:specbestB}
\end{figure}

\begin{figure}[p]
  \centering
{\begin{turn}{-0}
  \psfig{figure=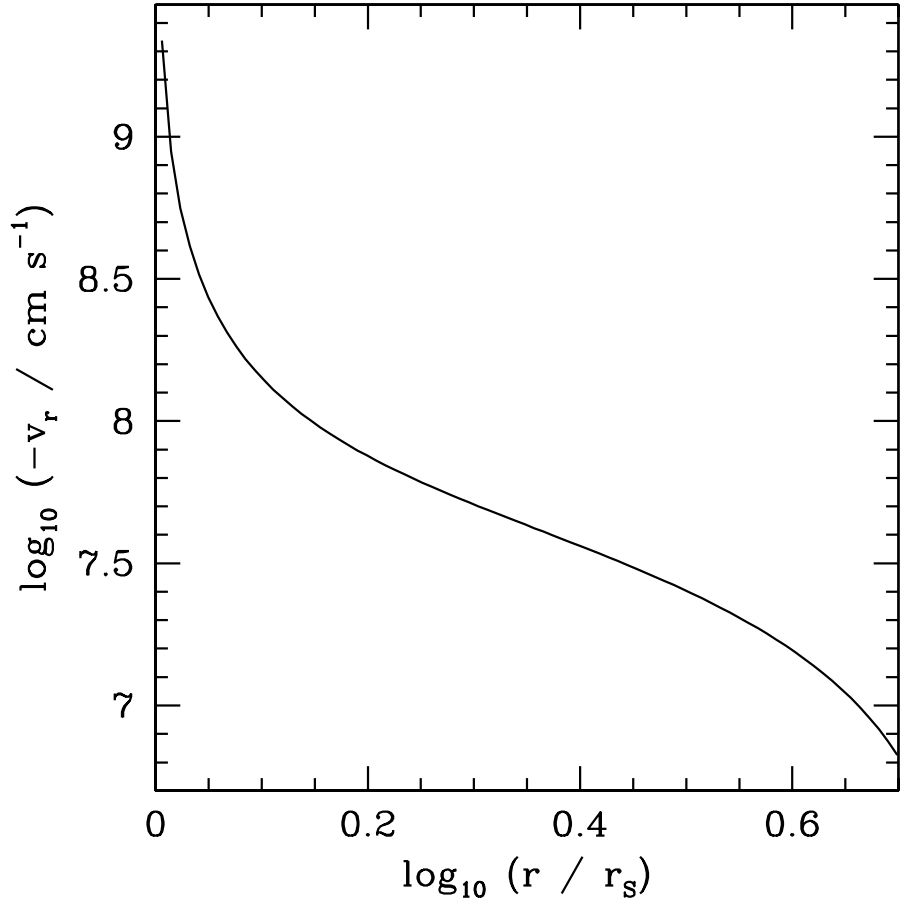,width=5.in}
\end{turn}}
  \caption{Radial velocity profile corresponding to the best fit model, shown in Figure 1.}
  \label{fig:specbestvr}
\end{figure}

\begin{figure}[p]
  \centering
{\begin{turn}{-0}
  \psfig{figure=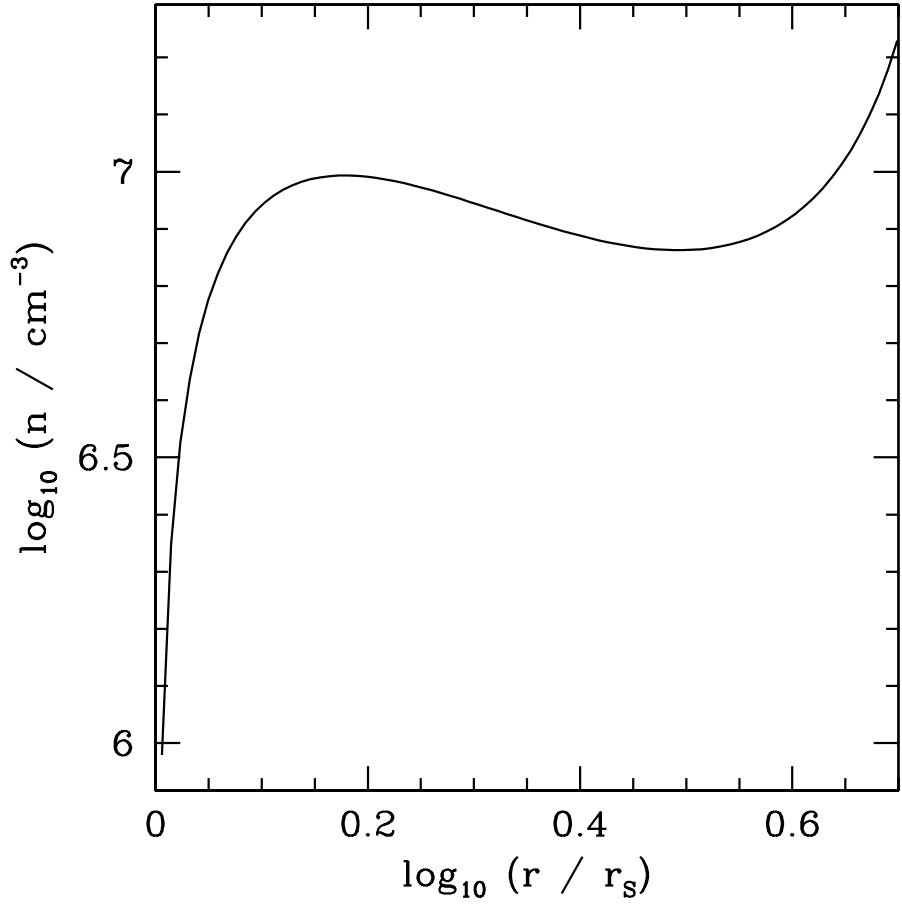,width=5.in}
\end{turn}}
  \caption{Particle number density corresponding to the best fit model, shown in Figure 1.}
  \label{fig:specbestn}
\end{figure}

\begin{figure}[p]
  \centering
{\begin{turn}{-0}
  \psfig{figure=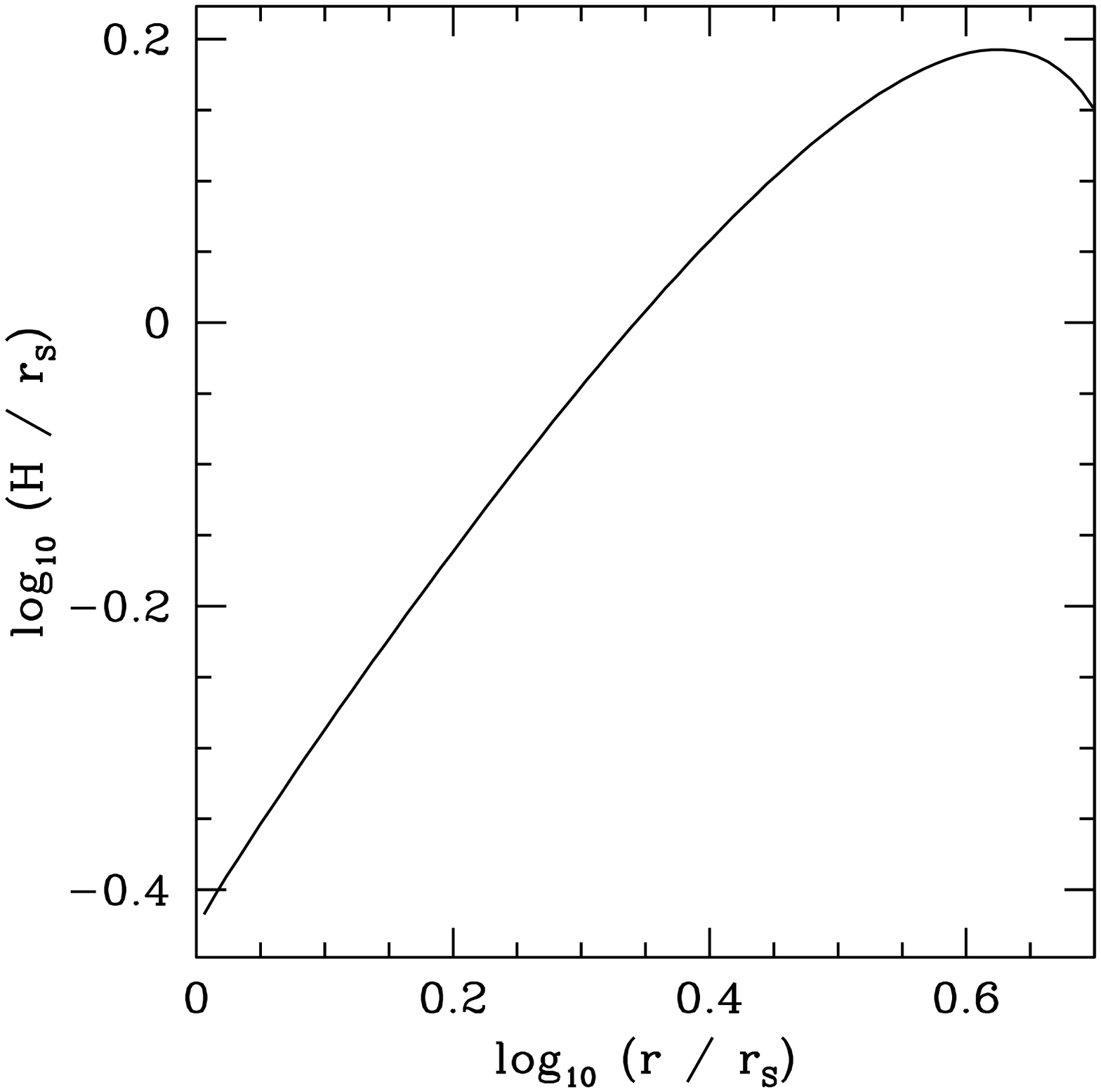,width=5.in}
\end{turn}}
  \caption{Scale height of the gas as a function of $r$, divided by the Schwarzschild radius, 
for the best fit model, shown in Figure 1.}
  \label{fig:specbestH}
\end{figure}

The requirement that the flow remains optically thin in the Keplerian
region (i.e., within the nominal $5\;r_S$ we are considering here)
means that $n\sigma_T H\le \cos{i}$, where
$\sigma_T=6.65\times 10^{-25}$ cm$^2$ is the Thomson scattering cross section. 
By using Equation (\ref{mdot}) and $\Sigma=2Hnm_p$, we therefore have
\begin{equation}
\dot M\le -4\pi m_p r v_r \cos{i}/\sigma_T\; .
\end{equation}
But Equations (\ref{vr2}) and (\ref{vr1}) show that 
\begin{equation}
-v_r= {Tr\beta_\nu\beta_pR_g\over \mu(GM)^{1/2}(r^{1/2}-r_i^{1/2})}\; .\label{vr3}
\end{equation}
Combining these two relations, we get  
\begin{equation}
\dot M \le 1.8\times 10^{20}{\cos{i}\over \sqrt{r/r_i}-1}\left({r\over
r_i}\right)^2\left({T\over
10^{11}\;\hbox{K}}\right)\left({\beta_\nu\over 0.27}\right)\left({\beta_p\over 0.036}\right)
\;\hbox{g}\;{\hbox{s}}^{-1}\; .
\end{equation}
Since the factor $(r/r_i)^2/(\sqrt{r/r_i}-1)$ has a minimum value of $4^4/3^3$ 
at $r=16/9\ r_S$, and the Keplerian flow must be optically thin everywhere
\begin{equation}
\dot M\le 1.7\times 10^{21} \cos{i} \left({T_m\over
10^{11}\;\hbox{K}}\right)\left({\beta_\nu\over 0.27}\right)\left({\beta_p\over 0.036}\right)
\;\hbox{g}\;{\hbox{s}}^{-1}\; .
\end{equation}
where $T_m$ is the maximal temperature.  In itself, this value is consistent
with the Bondi-Hoyle rate inferred from the simulations by Coker \& Melia (1997).

An even stricter upper limit for the accretion rate can be obtained if we assume that the
sub-millimeter bump in the radio is produced by a hot Keplerian flow through
synchrotron emission. This plasma will also produce X-radiation via 
bremsstrahlung emission. The bremsstrahlung emissivity is given by
\begin{equation}
\epsilon_b=6.8\times 10^{-38} n^2 T^{-0.5} e^{-h\nu/kT}\;\hbox{ergs}\;{\hbox{cm}}^{-3}\;
{\hbox{s}}^{-1}\; {\hbox{Hz}}^{-1}\; .
\end{equation}
In the Keplerian flow, $h\nu\ll kT$ within the $2-10$ keV energy range. 
Thus, with $\exp(-h\nu/kT)\approx 1$, we have 
\begin{equation}
L=\int_{2\;\hbox{keV}}^{10\;\hbox{keV}} {d(h\nu)\over h}\int_{r_i}^{r_0} dr\; 
4\pi r H \epsilon_b =2.66\times 10^{33}\;{\rm ergs}\;{\rm s}^{-1}\;,
\end{equation}
and using Equation (\ref{mdot}), this gives
\begin{equation}
\dot M^2\int_{r_i}^{r_0} {dr\over v_r^2Hr\sqrt{T}}\le 0.69\times 10^6\qquad
(\hbox{c.g.s.}).
\end{equation}
But from Equations (\ref{vr3}) and (\ref{H}) we have (again in c.g.s. units)
\begin{eqnarray}
\int_{r_i}^{r_0} {dr\over v_r^2Hr\sqrt{T}}\ge&&8.3\times
10^{-30}\int_1^{r_0/r_i}\left({10^{11}\;\hbox{K}\over T}\right)^3\left({0.27\over
\beta_\nu}\right)^2
\left({0.036\over \beta_p}\right)^2 {(\sqrt{x}-1)^2\over x^{4.5}} dx \nonumber \\
\ge1.7\times
10^{-29}&&\left({10^{11}\;\hbox{K}\over T_m}\right)^3\left({0.27\over
\beta_\nu}\right)^2
\left({0.036\over \beta_p}\right)^2 \left({1\over 105}-{1\over 5x_0^{2.5}}+{1\over
3x_0^3}-{1\over
7x_0^{3.5}}\right)\; ,
\end{eqnarray}
where $x_0=r_0/r_i$. So 
\begin{equation}
\dot M\le 2.1\times 10^{17}\left({T_m\over
10^{11}\;\hbox{K}}\right)^{1.5}\left({\beta_\nu\over
0.27}\right)\left({\beta_p\over 0.036}\right) \left({1\over 105}-{1\over 5x_0^{2.5}}+{1\over
3x_0^3}-{1\over
7x_0^{3.5}}\right)^{-0.5}\;\;\hbox{g}\;{\hbox{s}}^{-1}\;.
\end{equation}

\begin{figure}[p]
  \centering
{\begin{turn}{-0}
  \psfig{figure=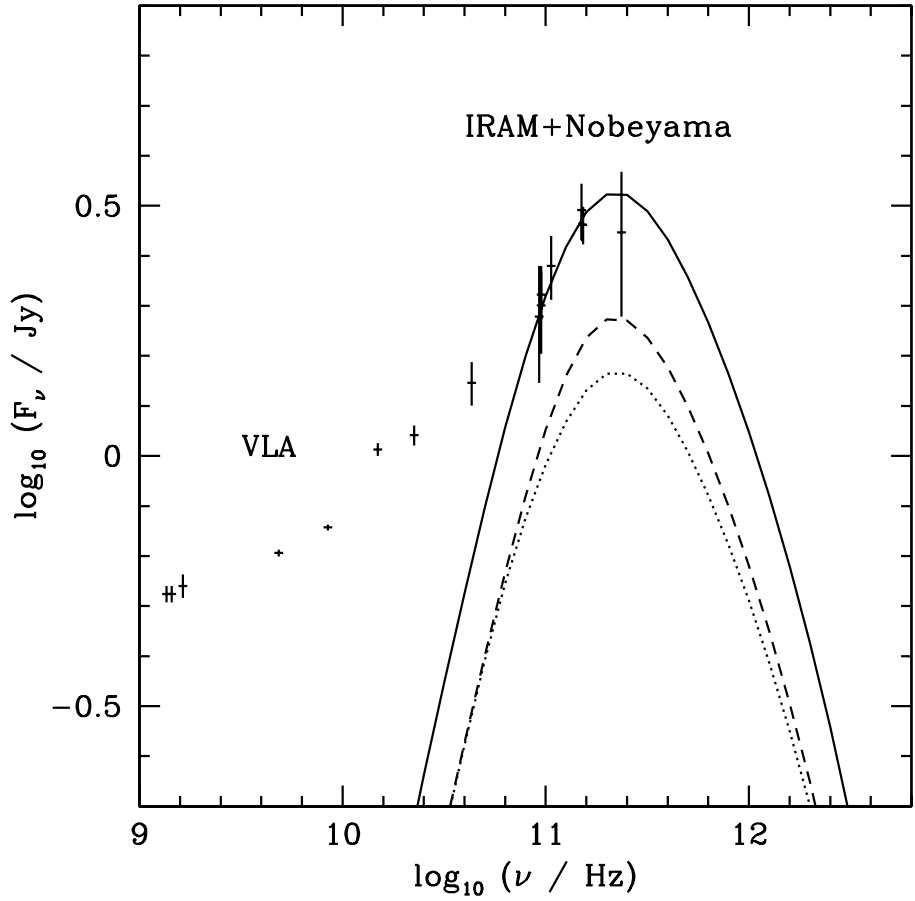,width=5.in}
\end{turn}}
  \caption{Enlargement of the radio portion of the spectrum for the best fit
model shown in Figure 1.  The dotted curve here corresponds to the first component. The dashed 
curve corresponds to the second component. The values of the parameters are those specified in
Figure 1.}
  \label{fig:specbestenlarge}
\end{figure}

\begin{figure}[p]
  \centering
{\begin{turn}{-0}
  \psfig{figure=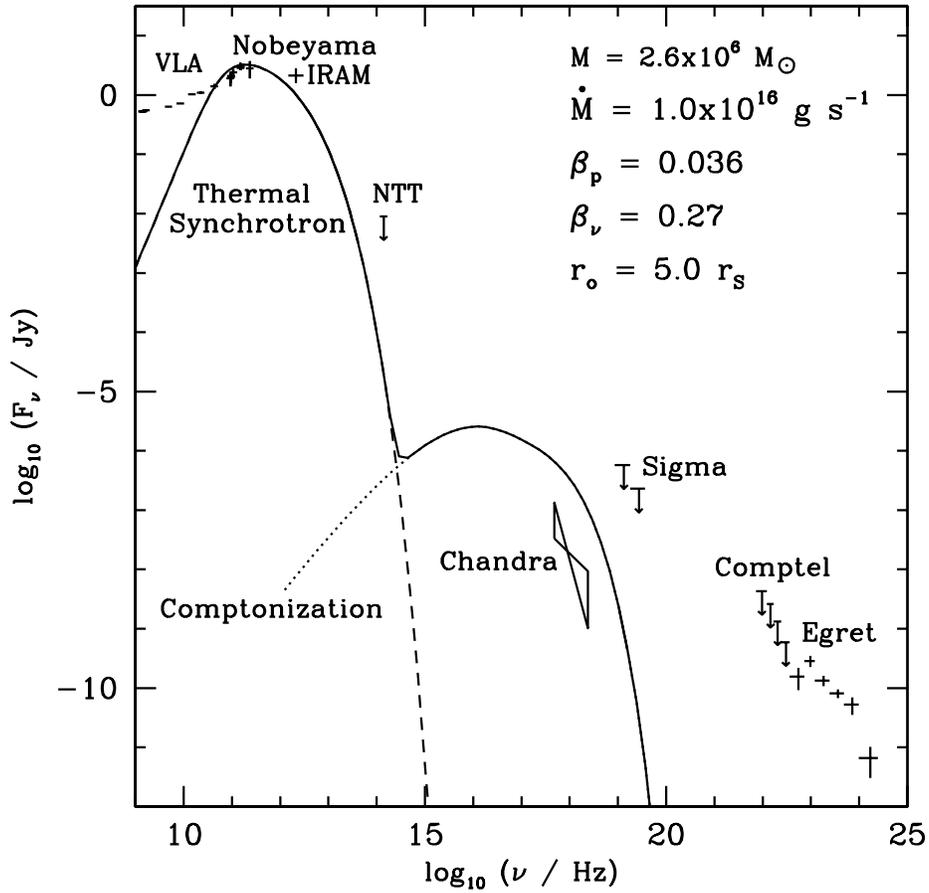,width=5.in}
\end{turn}}
  \caption{Illustration of the effect on the spectrum due to a change
in the ratio $v_r/v_{ff}$ at $r_0$.  In this case, it is $5.0\times 10^{-3}$, compared
to a value ten times smaller for the best fit model.}
  \label{fig:speclargevr}
\end{figure}

\begin{figure}[p]
  \centering
{\begin{turn}{-0}
  \psfig{figure=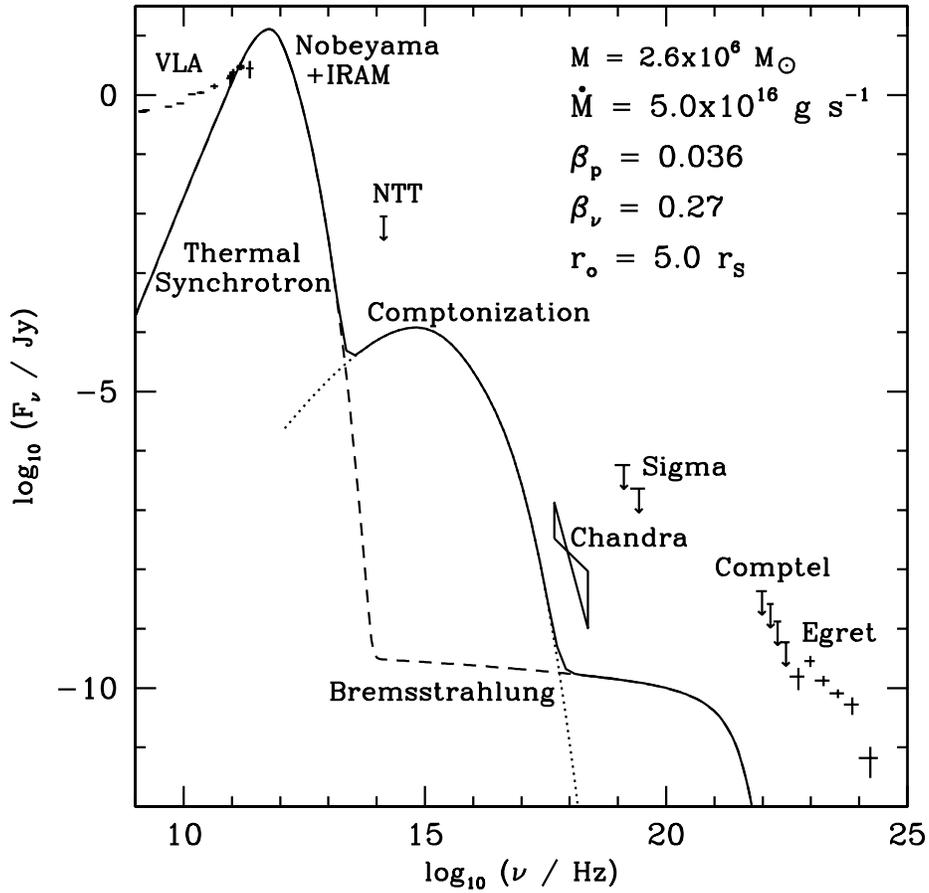,width=5.in}
\end{turn}}
  \caption{Illustration of the effect on the spectrum due to a change
in the value of $\dot M$, which is here five times larger than that of   
the best fit model shown in Figure 1.
}
  \label{fig:speclargemdot}
\end{figure}
We point out here that the range of values giving reasonable fits
to the sub-mm data (e.g., Fig. 1) falls below this limit, though close
to it. Thus, the current high-energy observations of Sgr A* do not
appear to be in conflict with this model.  Indeed, as one can see
from this figure, the self-Comptonized spectrum fits the {\it Chandra}
data rather well (Baganoff et al. 2000).

To sample the dependence of our fit on the boundary conditions, 
we show in Figure 8 the spectrum resulting from a Keplerian flow
forming with a higher radial velocity at its outer boundary. Equation (\ref{vr2}) shows
the outer boundary temperature inceases by the same order. The
implied higher viscosity results in a higher temperature over a
more extended range in radii, producing a much flatter ``hump''
that does not adequately fit the data.  This also produces too
many X-rays, violating the {\it Chandra} limit. However, increasing
the accretion rate through this Keplerian region will raise the cooling rate so
the gas temperature decreases. But the increased column density results in a
spectral bump peaking at  higher frequencies than what is observed.  In addition, these
conditions lead to a lower X-ray flux than is observed, though this might still be
consistent if the {\it Chandra} source is not the actual counterpart to
Sgr A*.

\section{Concluding Remarks}

Our calculations have shown that under the right conditions,
the thermal synchrotron emission from a compact, Keplerian
flow around Sgr A* can account for the spectral (sub-mm) bump.
Clearly, the peak frequency of this component is correlated
to the gas temperature and column density, which characterize several key
ingredients, such as the accretion rate and the advected
specific angular momentum.  The latter appears to be consistent
with the values inferred from the large scale Bondi-Hoyle
accretion (Coker \& Melia 2000), but the required $\dot M$ 
within the circularized region is significantly lower
than that expected further out.  This contrast probably indicates
the need for a gradual mass loss, perhaps in the form
of a wind, toward smaller radii.  In other words, compared to the value of
$\dot M$ inferred from the hydrodynamic simulations, the
maximum accretion rate permitted by the sub-mm data is
several orders of magnitude smaller.  Yet the specific angular
momentum with which the gas circularizes is not noticeably different
from that prescribed by the numerical simulations.

We note that this model would not pass the {\it Chandra} test
(of a very low X-ray luminosity) if the physical conditions
required to produce the sub-mm emission were more extreme than
those sampled in this paper.  Specifically, the sub-mm and X-ray
data would not be mutually consistent within the context of
this model if the accretion
rate through the Keplerian region was at the Bondi-Hoyle value. 

At this point, we can only speculate about what the overall
accretion pattern might be like, and we target a future coupling
between this analysis and a hydrodynamic simulation of the 
Keplerian flow merged with the larger scale accretion to
address the issue of self-consistency.  We anticipate that when
the quasi-spherical infall begins to circularize (presumably
around $100-1000\,r_S$), the turbulent mixing of gas elements 
with high eccentricity readily dissipates the gravitational and
kinetic energy densities, raising the internal energy of the
gas and possibly leading to a rapid evaporation away from the
rotation plane.  This expulsion of a wind may cease when the
eccentricity reaches a value close to zero, which should occur
at a radius $\sim 5-10\;r_S$.  At this point, the flow
is very nearly Keplerian, and the dynamo process may become
active and efficient, along the lines we have developed in this paper.
The spectrum of Sgr A* longward of $1-2$ mm is presumably
generated within the turbulent mixing (i.e., transition) region.

The first generation of hydrodynamic simulations to address
these questions are currently underway and we hope to report
the results of this analysis in the very near future.
 
{\bf Acknowledgments} We are very grateful to Fred Baganoff for
sending us the early results of his X-ray analysis of the
{\it Chandra} data, and to Marco Fatuzzo for his helpful
discussions.  This work was supported by a Sir Thomas Lyle Fellowship 
and a Miegunyah Fellowship for distinguished overseas visitors at the University 
of Melbourne, and by NASA grants NAG5-8239 and NAG5-9205 at the University
of Arizona.


\begin{thebibliography}{}
\bibitem{Bag}  Baganoff, F., et al. 2000, ApJ, to be submitted
\bibitem{BG1}  Balbus, S.A., Gammie, C.F., and Hawley, J.F. 1994, MNRAS, 271, 197 
\bibitem{BH1}  Balbus, S.A., \& Hawley, J.F. 1991, ApJ, 376, 214 (BH1)
\bibitem{BH2}  Balbus, S.A., \& Hawley, J.F. 1992, ApJ, 400, 610 (BH2)
\bibitem{BB73} Balick, B. \& Brown, RL 1974, ApJ, 194, 265
\bibitem{BNS} Brandenburg, A., Nordlund, \AA., Stein, R. \& Torkelsson, U. 1995, ApJ, 446, 741
(BNST)
\bibitem{CM} Coker, R.F., \& Melia, F. 2000, ApJ, 534, 723
\bibitem{Melia1} Coker, R.F., \& Melia, F. 1997, ApJ Letters, 488, L149
\bibitem{EG96} Eckart, A. \& Genzel, R. 1996, Nature, 383, 415
\bibitem{EG97} Eckart, A. \& Genzel, R. 1997, MNRAS, 284, 576
\bibitem{F98} Falcke, H., Goss, WM., Matsuo, H., Teuben, P., Zhao, J-H \& 
Zylka, R. 1998, ApJ, 499, 731
\bibitem{FSH}  Fleming, T.P., Stone, J.M., \& Hawley, J.F. 2000, ApJ, 530, 464
\bibitem{GEN96} Genzel, R., Thatte, N., Krabbe, A., Kroker, H. \& 
Tacconi-Garman, LE. 1996, ApJ, 472, 153
\bibitem{G98} Ghez, AM., Klein, BL., Morris, M. \& Becklin, EE. 1998, ApJ, 509, 678
\bibitem{Haw} Hawley, J.F., 2000, ApJ, 528, 462
\bibitem{HB1}  Hawley, J.F., \& Balbus, S.A. 1991, ApJ, 376, 223 (HB1)
\bibitem{HB2}  Hawley, J.F., \& Balbus, S.A. 1992, ApJ, 400, 595 (HB2)
\bibitem{HGB1}  Hawley, J.F., Gammie, C.F., \& Balbus, S.A. 1995, ApJ, 440, 742 (HGB1)
\bibitem{HGB2}  Hawley, J.F., Gammie, C.F., \& Balbus, S.A. 1996, ApJ, 464, 690 (HGB2)
\bibitem{IP77} Ipser, J.R. \& Price, R.H. 1977, ApJ, 216, 578.
\bibitem{Jin} Jin, L. 1996, ApJ, 457, 798
\bibitem{kow} Kowalenko, V. \& Melia, F. 1999, MNRAS, 310, 1053
\bibitem{Me92} Melia, F. 1992, ApJ Letters, 387, L25.
\bibitem{Melia} Melia, F. 1994, ApJ, 426, 577
\bibitem{MF}    Melia, F. \& Fatuzzo, M. 1989, ApJ, 346, 378
\bibitem{MJN92} Melia, F., Jokipii, JR. \& Narayanan, A. 1992, ApJ Letters, 395, 87
\bibitem{Pach} Pacholczyk, A.G. 1970, Radio Astrophysics, (W.H. Freeman and Company: San
Francisco)
\bibitem{pet64} Petschek, H.E. 1964, Proc. of the Symposium on Physics of Solar 
Flares (NASA SP-50), 425.
\bibitem{Sh73} Shapiro, S.L. 1973, ApJ, 185, 69.
\bibitem{Lyman} Spitzer, Lyman, Jr. 1961, Physics of Fully Ionized Gases, (John Wiley \& Sons: New York)
\bibitem{Stoeger} Stoeger, W.R. 1980, ApJ, 235, 216
\bibitem{SHGB}  Stone, J.M., Hawley, J.F., Gammie, C.F., \& Balbus, S.A. 1996, ApJ, 463, 656 (SHGB)
\bibitem{vho79} van Hoven, G. 1979, ApJ, 232, 572.
\bibitem{Z92} Zylka, R., Mezger, PG. \& Lesch, H. 1992, AA, 261, 119
\bibitem{Z95} Zylka, R., Mezger, PG., Ward-Thompson, D., Duschl, WJ. \& Lesch, H. 1995, AA, 297, 83
\end{thebibliography}
\end{document}